\documentclass[sn-mathphys]{sn-jnl}

\jyear{2022}%

\usepackage{amsmath}
\usepackage{amsfonts}
\usepackage{graphicx}
\usepackage{float}
\usepackage{color}
\usepackage{soul}
\usepackage{array}
\usepackage{placeins}

\newcommand{\beq}{\begin{equation}}
\newcommand{\eneq}{\end{equation}}
\newcommand{\be}{\begin{equation}}
\newcommand{\ee}{\end{equation}}
\newcommand{\bea}{\begin{eqnarray}}
\newcommand{\eea}{\end{eqnarray}}

\usepackage{wasysym}

\begin{document}

\title[Article Title]{Understanding traffic jams using Lindblad superoperators}

\author*{\fnm{Andrea} \sur{Nava*}}\email{andrea.nava@fis.unical.it}

\author{\fnm{Domenico} \sur{Giuliano}}

\author{\fnm{Alessandro} \sur{Papa}}

\author{\fnm{Marco} \sur{Rossi}}

\affil{\begin{center} Dipartimento di Fisica, Universit\`a della Calabria, \\ Arcavacata di Rende I-87036, Cosenza, Italy \\ INFN - Gruppo collegato di Cosenza, \\ Arcavacata di Rende I-87036, Cosenza, Italy \end{center}}

\abstract{We propose a model to simulate different traffic-flow conditions in terms of quantum graphs hosting an ($N$+1)-level dot at each site. Our model allows us to keep track of the type and of the destination of each vehicle. The traffic flow inside the system is encoded in a proper set of Lindbladian local dissipators that describe the time evolution of the system density matrix. Taking advantage of the invariance of the Lindblad master equation under inhomogeneous transformations we derive the quantum Hamiltonian for the bulk dynamics in a proper experimental setup.}

\keywords{Traffic flow, Traffic jam, Lindblad master equation, Open quantum systems}



\maketitle

\section{Introduction}
\label{intro}

In the last years, despite some initial skepticism, quantum methods turned out to be an excellent tool to describe the dynamics of classical complex macroscopic systems in a rich variety of fields as biology, social science, finance, logistics. While the literature is too wide to be reported here, a detailed review of different quantum approaches and an extended bibliography for each topic can be found, for example, in Refs.~\cite{khrennikov_1, asano, khrennikov_2, bagarello_1, bagarello_2, baaquie, busemeyer, galam}.
A huge number of analytical and numerical approaches to the traffic problem consist in modeling traffic as a fluid flow, or as a gas of interacting particles. Other approaches, like Cellular Automaton (CA) models or Fock space technique, adopt a microscopical description where vehicles are treated as particles moving on a discrete lattice. Within this second framework, we have recently introduced a fully quantum formulation in terms of a Markovian Master Lindblad Equation (LE) description for the density matrix~\cite{nava_traffic}. In our model, the classical stochastic rules for the time evolution are encoded into a set of 
so-called Lindblad operators acting on a proper Hilbert space. The Lindblad operators are chosen in order to preserve vehicle number conservation and monodirectional flows in the bulk, but are also able to introduce open boundary conditions on the border of the system. Such an approach gives us full access to both the steady state flow and the time evolution away from the steady-state: this allows for a complete description of the system dynamics without resorting, a priori, to any mean field approximation. Finally, despite the traffic flow is a classic problem, a full quantum statistical formulation opens up the possibility to use spin systems or cold atoms systems as microscopic simulators of real traffic situations.

The paper is organized as follows. In Sec.(\ref{sec:lind}) we briefly review the approach introduced in Ref.\cite{nava_traffic}, introduce the recipe to build the Fock space of a generic network, the LE, and the Lindblad jump operators and discuss the invariance properties under inhomogeneous transformations. In Sec.(\ref{sec:tasep}) we discuss the Totally Asymmetric Simple Exclusion Process (TASEP) model, that describes the flow of a single type of vehicles on a monodirectional line, and present Lindblad operators and the underlying spin Hamiltonian. In Sec.(\ref{sec:2s_tasep}) we introduce the two-species TASEP model to describe the monodirectional flow of two different type of vehicles, with different driving behaviors. In Sec.(\ref{sec:2l_tasep}) we discuss the two-line TASEP model which permits also the description of the overtaking process. Sec.(\ref{sec:conclusion}) concludes our study.

\section{The model}
\label{sec:lind}

In this section we review the ($N$+1)-level system for traffic model developed in Ref.\cite{nava_traffic}.
The main idea is to divide a generic road network into sections of length $a$, where $a$ is the mean dimension of a vehicle. Then, we describe each section as a quantum $(N+1)$-level dot. The state $\left. \vert 0\right\rangle$ corresponds to an empty road
section, while the other $N$ states describe different vehicle types and destinations. Such states are orthogonal to each other and on each site we introduce the operators: $\sigma_{j,0}=\left. \vert j\right\rangle \left\langle 0 \vert \right.  $ ($\sigma_{0,j}=\left. \vert 0\right\rangle \left\langle j \vert \right.$), that create  (destroy) a given vehicle-destination combination, and $\sigma_{j', j}=\left. \vert j'\right\rangle \left\langle j \vert \right.$ that describe a vehicle of given type changing its destination or driving behavior. These operators satisfy the conditions $\sigma_{i,j}\sigma_{k,\lambda}=\delta_{j,k}\sigma_{i,\lambda}$, $\sigma_{i,j}^\dagger=\sigma_{j,i}$ and $\sum\limits_{j=0}^N \sigma_{j,j}=1$. As a consequence the dot can only be empty or occupied by a
single vehicle/destination combination as it is impossible to create more than one vehicle on the same site.
The Hilbert space for the complete network is a tensor product of the single-site Hilbert spaces. The basis vectors are defined as
\begin{equation}
\left. \vert j_{1},\ldots,j_{L}\right\rangle =\left. \vert j_{1}\right\rangle \otimes\ldots\otimes\left. \vert j_{L}\right\rangle \, ,
\end{equation}
while the creation, annihilation and conversion operators acting on a given dot $\ell$ are realized as  the tensor product of ($L-1$)  identity matrices of dimension $N_{\ell} \times N_{\ell}$ and the $\sigma_{i,j}$ operator at the site $\ell$,
that is
\begin{equation}
\sigma_{i,j}^{\left(\ell\right)}=\mathbb{I}\otimes\ldots\otimes\sigma_{i,j}\otimes\ldots\otimes\mathbb{I} \, .
\end{equation}
Finally, the incoherent (classical) dynamics of the open quantum system is expressed in terms of a LE for the time evolution
of the density matrix $\rho ( t )$, describing  the interaction between different dots and between the boundary dots and a set of external reservoirs in terms of the jump, or Lindblad,
operators, $\mathcal{L}_k$:
\begin{eqnarray}
\dot{\rho}\left(t\right)&=&\sum_{k}\left(\mathcal{L}_{k}\rho \left(t\right)  \mathcal{L}_{k}^{\dagger}-\frac{1}{2}\left\{ \mathcal{L}_{k}^{\dagger}\mathcal{L}_{k},\rho \left(t\right) \right\} \right)
\label{eq:lindbladeq} \, ,
\end{eqnarray}
where $k$ labels all the different Lindblad jump operators.

In our model we introduce two kinds of jump operators.  Operators of the first kind act locally on the boundary sites, creating and destroying vehicles;
instead, operators of the second kind  describe the incoherent stochastic transport of vehicles, thus  playing  a role similar to the Hamiltonian governing
the coherent transport in the Liouvillian equation. All these operators can be expressed in terms of the $\sigma_{i,j}^{(\ell)}$ operators.
The creation and annihilation Lindblad operators are defined as
\begin{eqnarray}
\mathcal{L}_{{\rm in},j}^{(\ell)} & = & \sqrt{\Gamma_{j}^{(\ell)}} \sigma^{\left(\ell\right)}_{j,0}\nonumber \\
\mathcal{L}_{{\rm out},j}^{(\ell)} & = & \sqrt{\gamma_{j}^{(\ell)}} \sigma^{\left(\ell\right)}_{0,j}\, ,
\end{eqnarray}
where $\Gamma_{j}^{(\ell)}$ and $\gamma_{j}^{(\ell)}$, with $\ell$ a boundary site, are the coupling constants to inject (in) or remove (out) a vehicle of kind $j$. They encode all the information regarding the ``environment'' that lies outside the system we are interested in.
The hopping operators that move a given vehicle from the site $\ell$ to the neighboring site $\ell'$ are, in the simplest case, defined as
\begin{equation}
\mathcal{L}^{(\ell,\ell')}_{{\rm hop},j,j'}=\sqrt{t_{j,j'}^{\left( \ell,\ell' \right)}}\sigma_{j',0}^{(\ell')}\sigma_{0,j}^{\left(\ell\right)} \, ,
\label{eq:hopp}
\end{equation}
\noindent
with $t_{j,j'}^{\left( \ell,\ell' \right)}$ the coupling constant associated to the hopping process that moves a vehicle from position $\ell$ to position $\ell'$ and possibly change its type from $j$ to $j'$. However, less trivial hopping processes can be easily introduced, as we do in the following sections, multiplying the hopping operator in Eq.(\ref{eq:hopp}) by a string operator, $\mathcal{O}$, that defines the condition under which the hopping process occurs.
The LE should contain only jump operators between neighboring sites in accordance with the correct driving directions, while the string operator $\mathcal{O}$ can be nonlocal. All coupling constants can be time dependent.

As claimed in Ref.\cite{nava_traffic}, this approach allows to treat at the same level classical incoherent evolution, and pure quantum coherent evolution, thus opening the possibility to use quantum dot systems as experimental devices to simulate classical traffic behavior. Furthermore it allows to convert CA descriptions of the traffic-flow problem into a quantum formalism. Indeed, Lindblad operators can be directly inherited from the transition rules of the CA models, taking advantage of the formal analogy between classical stochastic processes and quantum mechanical formalism.

Apparently, in Eq.(\ref{eq:lindbladeq}) the coherent evolution of a quantum lattice system, i.e. the Liouvillian-von Neumann contribution, is missing \cite{lindblad_nrg}. Indeed, the full quantum time evolution of the density matrix of an open quantum system should be
\begin{equation}
\dot{\rho}\left(t\right) = -i \left[ H, \rho \right] + \sum_{k}\left(\mathcal{L}_{k}\rho \left(t\right)  \mathcal{L}_{k}^{\dagger}-\frac{1}{2}\left\{ \mathcal{L}_{k}^{\dagger}\mathcal{L}_{k},\rho \left(t\right) \right\} \right)
\label{eq:lindbladeq_full} \, ,
\end{equation}
where at the right-hand side of Eq.(\ref{eq:lindbladeq_full}) the first term describes the unitary evolution of the quantum system due its own Hamiltonian $H$.
In our master equation formalism for the traffic problem the coherent evolution seems missing from the dynamics, see Eq.(\ref{eq:lindbladeq}), as we set $H=0$. This would induce to think that the comparison with a truly quantum system is incomplete and that it is not possible to simulate a classical traffic problem on a proper quantum device, as its intrinsic dynamics induced by the Hamiltonian is missing in our formulation. However, this is not true and we address this point in this paper. Indeed, the coherent dynamics can be reintroduced by a proper redefinition of the Lindblad operators making advantage of LE invariance properties: following Ref.\cite{petruccione}, the LE is invariant under inhomogeneous transformations, that is
\begin{align}
\mathcal{L}_{k} \equiv \sqrt{\gamma_{k}}\mathcal{O}_{k} & \rightarrow\hat{\mathcal{L}}_{k}=\sqrt{\gamma_{k}}\left( \mathcal{O}_{k}+a_{k} \mathbb{I}\right) \,\, , \nonumber \\
H & \rightarrow\hat{H}=H+\frac{\gamma_k}{2i}\sum_{k}\left(a_{k}^{*}\mathcal{L}_{k}-a_{k}\mathcal{L}_{k}^{\dagger}\right) \,\, ,
\label{eq:inhomogeneous}
\end{align}
with $a_k$ complex numbers.

In the following section we make use of the classical-to-quantum correspondence of our formalism in some simple cases and apply Eq.(\ref{eq:inhomogeneous}) to derive the quantum Hamiltonian that could describe the dynamics of an experimental setup. Vice versa, given an experimental setup with a proper Hamiltonian, the inhomogeneous transformation tells us how to modify the Lindblad operators (from the ones obtained, for example, from a classical CA without coherent evolution) in order to reproduce the same classical behavior.

\section{TASEP model}
\label{sec:tasep}

\begin{figure}[h]
\center
\includegraphics*[width=22pc]{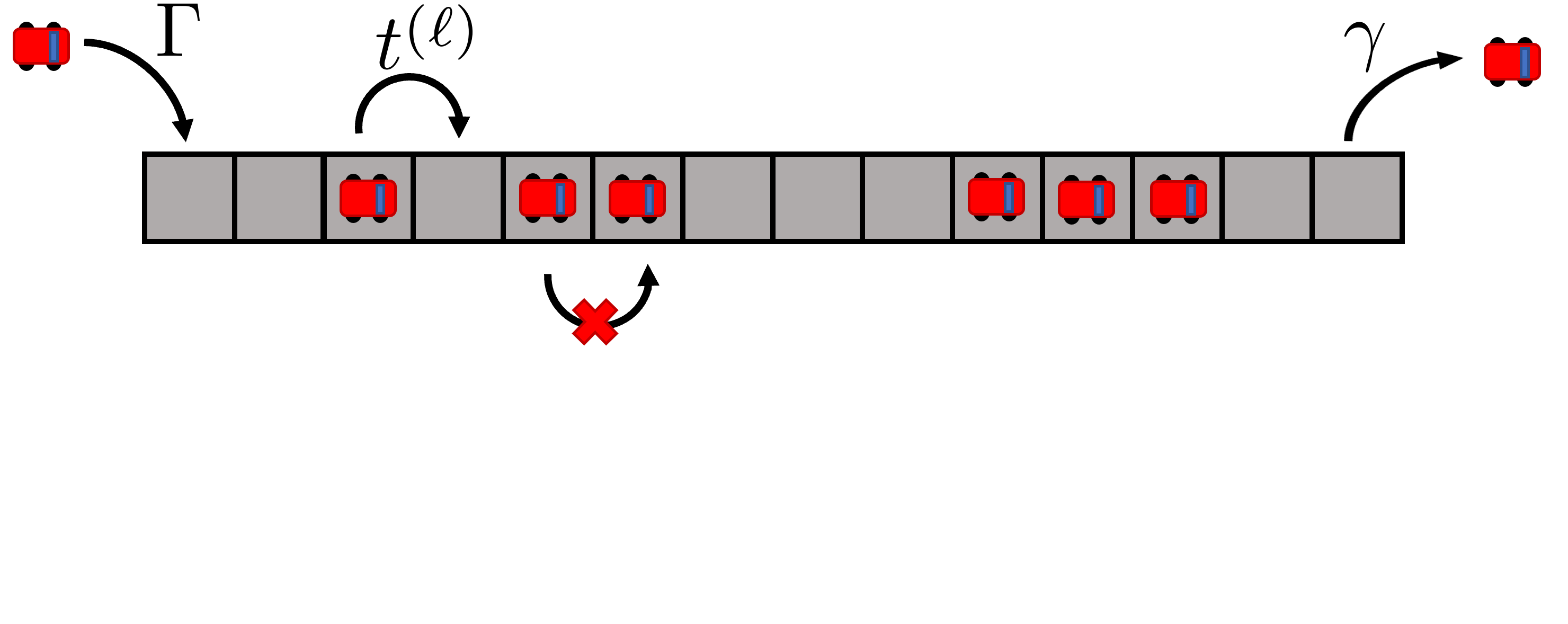}\hspace{2pc}%
\caption{\label{fig:tasep} Pictorial representation of the TASEP model. Vehicles are injected from the first site and removed from the last site with rates $\Gamma$ and $\gamma$, respectively. Vehicles can jump between two adjacent sites $\ell$ and $\ell+1$, from left to right, with rate $t^{(\ell)}$, only if the destination site is empty.}
\end{figure}

The simplest system that can be described by our formalism is the open boundary TASEP model that describes the monodirectional motion along a 1D line of hardcore particles (i.e. vehicles) on a lattice. In this framework, vehicles are injected from one site of the line and removed from the other endpoint \cite{shutz,temme}. Having only one kind of vehicles and one possible destination the system is described in terms of $\left( N=2 \right)$-level quantum dots. For a line of length $L$, the TASEP model is reproduced by setting $H=0$ and by introducing the following set of Lindblad jump operators
\begin{align}
\mathcal{L}_{{\rm in}}^{(1)} & =\sqrt{\Gamma}\sigma_{1,0}^{(1)} \,\, ,   \nonumber \\
\mathcal{L}_{{\rm out}}^{(L)} & =\sqrt{\gamma}\sigma_{0,1}^{(L)} \,\, , \\
\mathcal{L}_{{\rm hop}}^{(\ell)} & =\sqrt{t^{(\ell)}}\sigma_{1,0}^{(\ell+1)}\sigma_{0,1}^{(\ell)} \,\, , \ \ \ \ 1 \le \ell \le \left(L-1\right)\nonumber \, ,
\end{align}
where $\left. \vert 0 \right\rangle$ is for an empty road section and $\left. \vert 1 \right\rangle$ corresponds to an occupied section; $\Gamma$ and $\gamma$ are the coupling constants for the incoming and outgoing flows, while the $t^{(\ell)}$ are the hopping coupling constants. The coupling constants can in general depend on time, $t$ (see Fig.(\ref{fig:tasep}) for a pictorial representation).

\noindent
Applying Eq.(\ref{eq:inhomogeneous}) we can shift the bulk hopping Lindblad operators, $\mathcal{L}_{{\rm hop}}^{(\ell)}$, by a constant in order to introduce the effective quantum Hamiltonian 
\begin{align}
\mathcal{L}_{{\rm hop}}^{(\ell)} & \rightarrow\hat{\mathcal{L}}_{{\rm hop}}^{(\ell)}=\sqrt{t^{(\ell)}}\left( \sigma_{1,0}^{(\ell+1)}\sigma_{0,1}^{(\ell)}+a_{\ell}\mathbb{I} \right) \,\, , \nonumber \\
H=0 & \rightarrow\hat{H}=\frac{1}{2i}\sum_{\ell}t^{(\ell)}\left(a_{\ell}^{*}\sigma_{1,0}^{(\ell+1)}\sigma_{0,1}^{(\ell)}-a_{\ell}\sigma_{0,1}^{(\ell+1)}\sigma_{1,0}^{(\ell)}\right) \, .
\label{eq:inhomogeneous-tasep}
\end{align}
If we set $a_\ell=2i$ and reinterpret the basis vectors as the spin-up and spin-down projection states along $z$ of a spin-$1/2$ particle, i.e. $\left. \vert 0 \right\rangle \rightarrow \left. \vert \downarrow \right\rangle$ and $\left. \vert 1 \right\rangle \rightarrow \left. \vert \uparrow \right\rangle$, we can map our system into an XX spin chain \cite{paletta} described by the Hamiltonian $\hat{H}$
\begin{equation}
\hat{H}=-\sum_{\ell} t^{(\ell)} \left(\sigma_+^{(\ell+1)}\sigma_-^{(\ell)}+\sigma_-^{(\ell+1)}\sigma_+^{(\ell)}\right)
\label{eq:spin-1/2} \, ,
\end{equation}
where $\sigma_i$, $i=x,y,z$, are the Pauli matrix and $\sigma_{\pm}=\sigma_x \pm i \sigma_y$. It follows that an XX spin chain plus the Lindblad operators $\mathcal{L}_{{\rm in}}^{(1)}$, $\mathcal{L}_{{\rm out}}^{(L)}$ and $\hat{\mathcal{L}}_{{\rm hop}}^{(\ell)}$ is a good experimental setup to simulate the classical TASEP model.

\section{Two-species TASEP}
\label{sec:2s_tasep}

\begin{figure}[h]
\center
\includegraphics*[width=22pc]{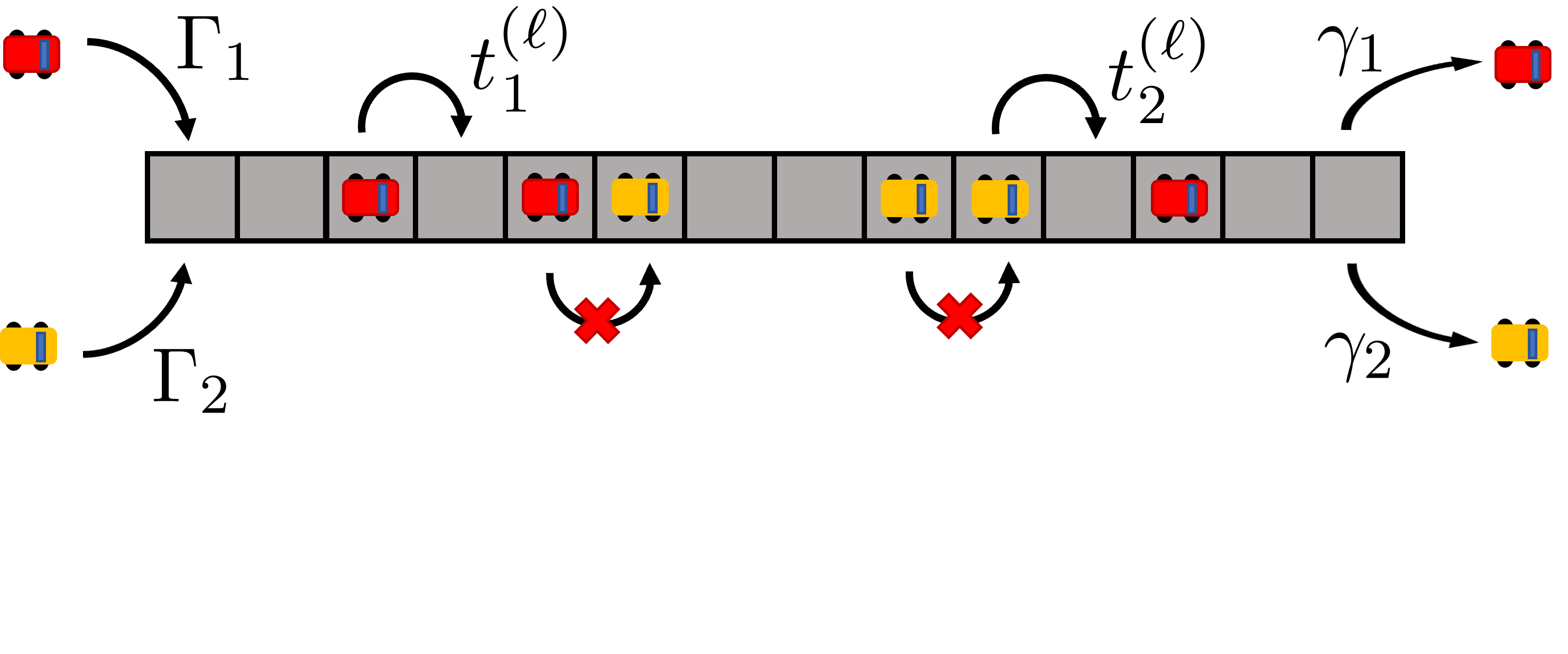}\hspace{2pc}%
\caption{\label{fig:2v_tasep} Pictorial representation of the two-species TASEP model. Two different kinds of vehicles, labeled with index $k=1,2$, are injected from the first site and removed from the last site with rates $\Gamma_k$ and $\gamma_k$, respectively. Vehicles can jump between two adjacent sites $\ell$ and $\ell+1$, from left to right, with rate $t_{k}^{(\ell)}$, only if the destination site is empty.}
\end{figure}

For a line of length $L$, the two-species TASEP model \cite{2tasep}, i.e. a TASEP model with two distinct vehicle types that could correspond to different destinations or speeds, is reproduced by setting $N=3$, $H=0$ and by introducing the following set of Lindblad jump operators
\begin{align}
\mathcal{L}_{{\rm in},k}^{(1)} & =\sqrt{\Gamma_k}\sigma_{k,0}^{(1)} \,\, ,\nonumber \\
\mathcal{L}_{{\rm out},k}^{(L)} & =\sqrt{\gamma_k}\sigma_{0,k}^{(L)} \,\, , \\
\mathcal{L}_{{\rm hop},k}^{(\ell)} & =\sqrt{t_{k}^{(\ell)}}\sigma_{k,0}^{(\ell+1)}\sigma_{0,k}^{(\ell)} \,\, ,\ \ \ \ 1 \le \ell \le \left(L-1\right)\nonumber , \,\, k=1,2 \, \, ,
\end{align}
where $\left. \vert 0 \right\rangle$ is for an empty road section and $\left. \vert 1 \right\rangle$, $\left. \vert 2 \right\rangle$ correspond to a section occupied by a vehicle of kind $1$ or $2$ respectively; $\Gamma_k$ and $\gamma_k$ are the coupling constants for the incoming and outgoing flows, while the $t_{k}^{(\ell)}$ are the hopping coupling constants, from site $\ell$ to site $\ell+1$, for vehicles of species $k$ (see Fig.(\ref{fig:2v_tasep}) for a pictorial representation).

\noindent
Applying again Eq.(\ref{eq:inhomogeneous}) to shift the bulk hopping Lindblad operators, $\mathcal{L}_{{\rm hop},k}^{(\ell)}$, by a constant, we have
\begin{align}
\mathcal{L}_{hop,k}^{(\ell)} & \rightarrow\hat{\mathcal{L}}_{hop,k}^{(\ell)}=\sqrt{t_{k}^{(\ell)}}\left( \sigma_{k,0}^{(\ell+1)}\sigma_{0,k}^{(\ell)}+a_{\ell,k}\mathbb{I} \right) \,\, , \nonumber \\
H=0 & \rightarrow\hat{H}=\frac{1}{2i}\sum_{\ell,k}t_{k}^{(\ell)}\left(a_{\ell,k}^{*}\sigma_{k,0}^{(\ell+1)}\sigma_{0,k}^{(\ell)}-a_{\ell,k}\sigma_{0,k}^{(\ell+1)}\sigma_{k,0}^{(\ell)}\right) \, .
\label{eq:inhomogeneous-2v_tasep}
\end{align}
In order to recover an analogy with a spin system, we interpret the basis vectors as the spin-up, spin-down and spin-0 projection states along $z$ of a spin-$1$ particle, i.e. $\left. \vert 0 \right\rangle \rightarrow \left. \vert 0 \right\rangle$, $\left. \vert 1 \right\rangle \rightarrow \left. \vert \uparrow \right\rangle$ and $\left. \vert 2 \right\rangle \rightarrow \left. \vert \downarrow \right\rangle$. If we now set $a_{\ell,k}=4i$ and introduce the Pauli matrices equivalent for spin-1 \cite{spin1}
\begin{equation}
\Sigma_{x}=\frac{1}{\sqrt{2}}\left(\begin{array}{ccc}
0 & 1 & 0\\
1 & 0 & 1\\
0 & 1 & 0
\end{array}\right) , \ \ \Sigma_{y}=\frac{i}{\sqrt{2}}\left(\begin{array}{ccc}
0 & -1 & 0\\
1 & 0 & -1\\
0 & 1 & 0
\end{array}\right) , \ \ \Sigma_{z}=\left(\begin{array}{ccc}
1 & 0 & 0\\
0 & 0 & 0\\
0 & 0 & -1
\end{array}\right) \, ,
\end{equation}
we can introduce the mapping
\begin{align}
\sigma_{1,0}^{\left(j\right)} & \rightarrow\frac{1}{\sqrt{2}}\Sigma_{+}^{\left(\ell\right)}\left(1+\Sigma_{z}^{\left(\ell\right)}\right) \,\, ,\nonumber \\
\sigma_{2,0}^{\left(\ell\right)} & \rightarrow\frac{1}{\sqrt{2}}\Sigma_{-}^{\left(\ell\right)}\left(1-\Sigma_{z}^{\left(\ell\right)}\right) \,\, ,\nonumber \\
\sigma_{0,1}^{\left(\ell\right)} & \rightarrow\frac{1}{\sqrt{2}}\Sigma_{-}^{\left(\ell\right)}\Sigma_{z}^{\left(\ell\right)} \,\, , \nonumber \\
\sigma_{0,2}^{\left(\ell\right)} & \rightarrow-\frac{1}{\sqrt{2}}\Sigma_{+}^{\left(\ell\right)}\Sigma_{z}^{\left(\ell\right)} \,\, ,
\end{align}
so that the Hamiltonian becomes
\begin{align}
\hat{H} & = -\sum_{\ell}\left[t_{1}^{(\ell)}\Sigma_{+}^{\left(\ell+1\right)}\left(\Sigma_{z}^{\left(\ell+1\right)}+1\right)\Sigma_{-}^{\left(\ell\right)}\Sigma_{z}^{\left(\ell\right)}+t_{2}^{(\ell)}\Sigma_{-}^{\left(\ell+1\right)}\left(\Sigma_{z}^{\left(\ell+1\right)}-1\right)\Sigma_{+}^{\left(\ell\right)}\Sigma_{z}^{\left(\ell\right)}\right]\nonumber \\
 & +\mathrm{h.c.} \,\, ,
\end{align}
where $\Sigma_{\pm}=\Sigma_x \pm i \Sigma_y$ and the $\Sigma_z$ that appears in the Hamiltonian is the string operator needed in order to avoid the conversion of a vehicle of type 1 into a vehicle of type 2 and vice versa.

\section{Two-line TASEP model}
\label{sec:2l_tasep}

\begin{figure}[h]
\center
\includegraphics*[width=22pc]{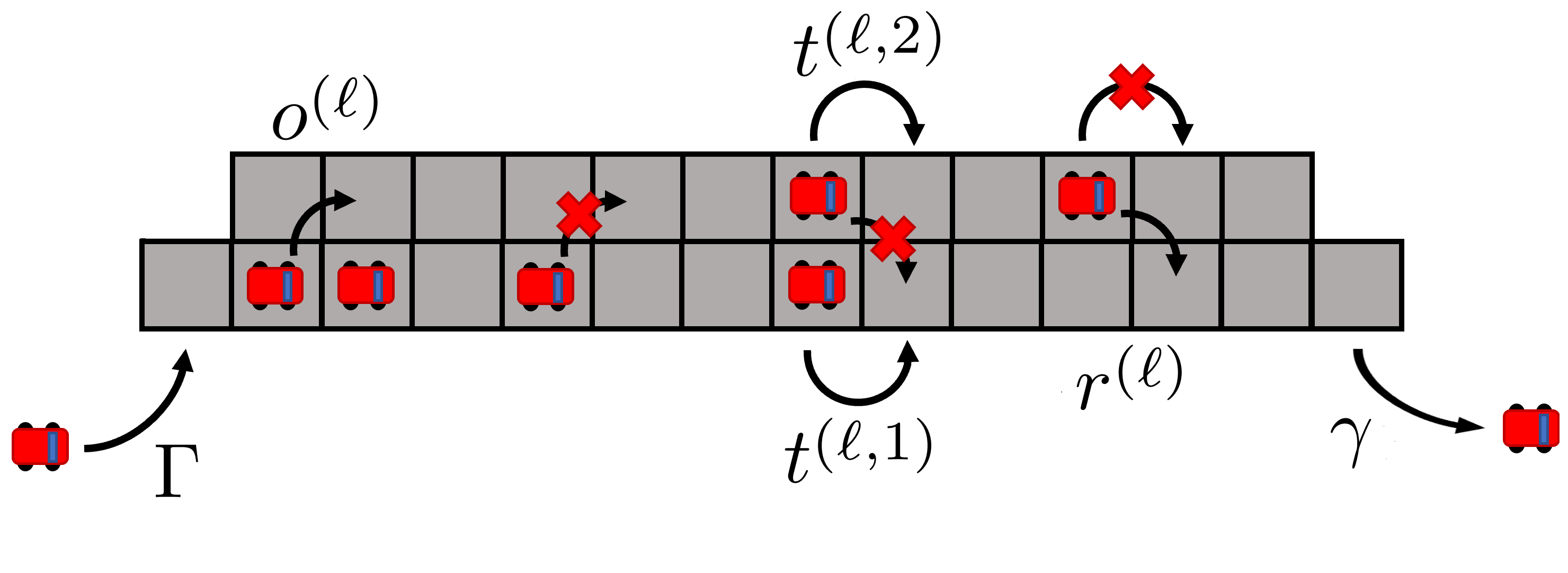}\hspace{2pc}%
\caption{\label{fig:2l_tasep} Pictorial representation of the two-line TASEP model. Vehicles are injected from the first site and removed from the last site with rates $\Gamma$ and $\gamma$, respectively. Vehicles can jump between two adjacent sites $\ell$ and $\ell+1$, from left to right, with rate $t^{(j,i)}$, only if the destination site is empty. Vehicles can also overtake another vehicle in front on them, taking advantage of the fast line. The overtaking is described in terms of the rates $o^{(\ell)}$ and $r^{(\ell)}$.}
\end{figure}

Let us consider a two-lane monodirectional road, with a regular lane $(y=1)$ and a fast lane $(y=2)$, in presence of a single kind of vehicle. Vehicles mainly move on lane $(y=1)$, however we introduce the possibility for a vehicle to overtake another vehicle in front of it, taking advantage of the fast line. Having only one kind of vehicles and one possible destination (the generalization to the two-line two-species TASEP model is trivial applying the recipe given in this and the previous sections) the system is again described in terms of $\left( N=2 \right)$-level quantum dots. However, for a line of length $L$, we have now $2L$ quantum dots labeled as $(x,y)$, with $x=1,...,L$ and $y=1,2$. The two-line TASEP model is reproduced by setting $H=0$ and by introducing the following set of Lindblad jump operators
\begin{align}
\mathcal{L}_{{\rm in}}^{(1,1)} & =\sqrt{\Gamma}\sigma_{1,0}^{(1,1)} \,\, ,\nonumber \\
\mathcal{L}_{{\rm out}}^{(L,1)} & =\sqrt{\gamma}\sigma_{0,1}^{(L,1)} \,\, ,\nonumber \\
\mathcal{L}_{{\rm hop}}^{(\ell,1)} & =\sqrt{t^{(\ell,1)}}\sigma_{1,0}^{(\ell+1,1)}\sigma_{0,1}^{(\ell,1)} \,\, ,\ \ \ \ \ \ \ \ \ \ \ \ 1 \le \ell \le \left(L-1\right) \,\, ,\nonumber \\
\mathcal{L}_{{\rm hop}}^{(\ell,2)} & =\sqrt{t^{(\ell,2)}}\sigma_{1,0}^{(\ell+1,2)}\sigma_{0,1}^{(\ell,2)} \sigma_{1,1}^{(\ell,1)} \,\, , \ \ \ \ \ 2 \le \ell \le \left(L-2\right) \,\, , \\
\mathcal{L}_{{\rm ove}}^{(\ell)} & =\sqrt{o^{(\ell)}}\sigma_{1,0}^{(\ell+1,2)}\sigma_{0,1}^{(\ell,1)} \sigma_{1,1}^{(\ell+1,1)} \,\, , \ \ \ \ 1 \le \ell \le \left(L-3\right) \,\, , \nonumber \\
\mathcal{L}_{{\rm ret}}^{(\ell)} & =\sqrt{r^{(\ell)}}\sigma_{1,0}^{(\ell+1,1)}\sigma_{0,1}^{(\ell,2)} \sigma_{0,0}^{(\ell,1)} \,\, , \ \ \ \ \ \ \ 2 \le \ell \le \left(L-1\right)\nonumber \nonumber \, ,
\end{align}
where $\left. \vert 0 \right\rangle$ is for an empty road section and $\left. \vert 1 \right\rangle$ corresponds to an occupied section; $\Gamma$ and $\gamma$ are the coupling constants for the incoming and outgoing flows from the regular line, while the $t^{(\ell,2)}$ and $t^{(\ell,1)}$ are the hopping coupling constants, with $t^{(\ell,2)}>t^{(\ell,1)}$. Finally, $o^{(\ell)}$ is the overtaking coupling constant and $r^{(\ell)}$ is the coupling constant for the return from overtaking. The overtaking can take place only if the following section road is occupied and the vehicle returns to the regular line as soon as the regular line next to it is empty (see Fig.(\ref{fig:2l_tasep}) for a pictorial representation).

\noindent
Applying Eq.(\ref{eq:inhomogeneous}), we can shift the bulk Lindblad operators, $\mathcal{L}_{{\rm hop}}^{(\ell,y)}$, $\mathcal{L}_{{\rm ove}}^{(\ell)}$ and $\mathcal{L}_{{\rm ret}}^{(\ell)}$, by a constant in order to introduce the effective quantum Hamiltonian 
\begin{align}
\mathcal{L}_{{\rm hop}}^{(\ell,1)} & \rightarrow\hat{\mathcal{L}}_{{\rm hop}}^{(\ell,1)}=\sqrt{t^{(\ell,1)}}\left( \sigma_{1,0}^{(\ell+1,1)}\sigma_{0,1}^{(\ell,1)}+a_{\ell}\mathbb{I} \right) \,\, , \nonumber \\
\mathcal{L}_{{\rm hop}}^{(\ell,2)} & \rightarrow\hat{\mathcal{L}}_{{\rm hop}}^{(\ell,2)}=\sqrt{t^{(\ell,2)}}\left( \sigma_{1,0}^{(\ell+1,2)}\sigma_{0,1}^{(\ell,2)}\sigma_{1,1}^{(\ell,1)} +a_{\ell}\mathbb{I} \right) \,\, , \nonumber \\
\mathcal{L}_{{\rm ove}}^{(\ell)} & \rightarrow\hat{\mathcal{L}}_{{\rm ove}}^{(\ell)}=\sqrt{o^{(\ell)}}\left( \sigma_{1,0}^{(\ell+1,2)}\sigma_{0,1}^{(\ell,1)} \sigma_{1,1}^{(\ell+1,1)}+a_{\ell}\mathbb{I} \right) \,\, , \nonumber \\
\mathcal{L}_{{\rm ret}}^{(\ell)} & \rightarrow\hat{\mathcal{L}}_{{\rm ret}}^{(\ell)}=\sqrt{r^{(\ell)}}\left( \sigma_{1,0}^{(\ell+1,1)}\sigma_{0,1}^{(\ell,2)} \sigma_{0,0}^{(\ell,1)}+a_{\ell}\mathbb{I} \right) \,\, , \nonumber \\
H=0 & \rightarrow\hat{H}=\frac{1}{2i}\sum_{\ell=1}^{L-1} t^{(\ell,1)}\left(a_{\ell}^{*}\sigma_{1,0}^{(\ell+1,1)}\sigma_{0,1}^{(\ell,1)}-a_{\ell}\sigma_{0,1}^{(\ell+1,1)}\sigma_{1,0}^{(\ell,1)}\right) \nonumber \\
& \ \ \ +\frac{1}{2i}\sum_{\ell=2}^{L-2}t^{(\ell,2)}\left(a_{\ell}^{*}\sigma_{1,0}^{(\ell+1,2)}\sigma_{0,1}^{(\ell,2)}\sigma_{1,1}^{(\ell,1)}-a_{\ell}\sigma_{0,1}^{(\ell+1,2)}\sigma_{1,0}^{(\ell,2)}\sigma_{1,1}^{(\ell,1)}\right) \nonumber \\ 
& \ \ \ +\frac{1}{2i}\sum_{\ell=1}^{L-3}o^{(\ell)}\left(a_{\ell}^{*}\sigma_{1,0}^{(\ell+1,2)}\sigma_{0,1}^{(\ell,1)} \sigma_{1,1}^{(\ell+1,1)}-a_{\ell}\sigma_{0,1}^{(\ell+1,2)}\sigma_{1,0}^{(\ell,1)} \sigma_{1,1}^{(\ell+1,1)}\right) \nonumber \\ 
& \ \ \ +\frac{1}{2i}\sum_{\ell=2}^{L-1}r^{(\ell)}\left(a_{\ell}^{*}\sigma_{1,0}^{(\ell+1,1)}\sigma_{0,1}^{(\ell,2)} \sigma_{0,0}^{(\ell,1)}-a_{\ell}\sigma_{0,1}^{(\ell+1,1)}\sigma_{1,0}^{(\ell,2)} \sigma_{0,0}^{(\ell,1)}\right) \, .
\label{eq:inhomogeneous-2l_tasep}
\end{align}
If we set $a_{\ell}=2i$ and reinterpret the basis vectors as the spin-up and spin-down projection states along $z$ of a spin-$1/2$ particle, i.e. $\left. \vert 0 \right\rangle \rightarrow \left. \vert \downarrow \right\rangle$ and $\left. \vert 1 \right\rangle \rightarrow \left. \vert \uparrow \right\rangle$, we can map our system into a spin ladder described by the Hamiltonian $\hat{H}$

\begin{align}
\hat{H} & = -\sum_{\ell=1}^{L-1} t^{(\ell,1)} \sigma_+^{(\ell+1,1)}\sigma_-^{(\ell,1)} -\sum_{\ell=2}^{L-2} t^{(j,2)} \sigma_+^{(\ell+1,2)}\sigma_-^{(\ell,2)} \left( \sigma_z^{(\ell,1)}+1 \right) \nonumber \\
& -\frac{1}{2} \sum_{\ell=1}^{L-3} o^{(\ell)} \sigma_+^{(\ell+1,2)}\sigma_-^{(\ell,1)} \left( \sigma_z^{(\ell+1,1)}+1 \right) -\frac{1}{2} \sum_{\ell=2}^{L-1} r^{(\ell)} \sigma_+^{(\ell+1,1)}\sigma_-^{(\ell,2)} \left( \sigma_z^{(\ell,1)}-1 \right) \nonumber \\
& + h.c \, ,
\end{align}
where $\sigma_i$, $i=x,y,z$, are the Pauli matrix and $\sigma_{\pm}=\sigma_x \pm i \sigma_y$. It follows that a spin ladder plus the Lindblad operators $\mathcal{L}_{{\rm in}}^{(1)}$, $\mathcal{L}_{{\rm out}}^{(L)}$, $\hat{\mathcal{L}}_{{\rm hop}}^{(\ell,y)}$, $\hat{\mathcal{L}}_{{\rm ove}}^{(\ell)}$ and $\hat{\mathcal{L}}_{{\rm ret}}^{(\ell)}$ is a good experimental setup for simulate the classical two-line TASEP model.

\section{Conclusions}
\label{sec:conclusion}

In this paper we discussed a master equation quantum approach for the description of the traffic problem by means of a graph of multilevel quantum dots coupled to external reservoirs. The jumps from/to the system and the internal flows are encoded into a set of Lindblad operators. The proposed formulation provides a microscopic description of a macroscopic system and allows to map a classical problem onto a quantum one paving the way to simulating traffic flow through quantum systems. Although in the original formulation of our approach \cite{nava_traffic} an explicit quantum Hamiltonian looks missing, we have shown that, taking advantage of the invariance properties of the LE, it is possible to unveil the correct combination of Hamiltonian and Lindblad operators that describe the current dynamics making the classical-to-quantum formulation more evident. The results shown in this paper can be easily generalized to the full network problem discussed in \cite{nava_traffic}. This allows to suggest how to extend results coming from classical CA algorithms \cite{ca_1,ca_2,ca_3,ca_4} to implement junction of fermionic or spin chains as well as junctions hosting exotic realizations of the Kondo effect \cite{fermi_1,fermi_2,fermi_3, fermi_5, kondo_0,kondo_1,kondo_2,kondo_3,ssh} as quantum playground for traffic flow simulations.
It is worth to note that, due the strong duality between CA rules and Lindblad jump operators, we believe that the master equation approach discussed in this paper in the context of traffic models, could be applied to other macroscopic systems as well. It follows that a comparison between LE approach and other quantum methods applied to complex classical systems, like for example the $(H,\rho)$-induced dynamics \cite{bagarello_3} or non-Hermitian Hamiltonian formalism \cite{graefe,bagarello_4}, would be extremely interesting and worth investigating.

\backmatter

\bmhead{Acknowledgments}

A.N., D. G., and M. R. acknowledge  financial support  from Italy's MIUR  PRIN projects TOP-SPIN (Grant No. PRIN 20177SL7HC). A.P. acknowledges support from the INFN project NPQCD. A.N. and D.G. acknowledge financial support from the INFN project SFT, M.R. acknowledges finnacial support from the INFN project GAST.




\begin{thebibliography}{9}
\bibitem{khrennikov_1} Khrennikov A. , Haven E. 2013 Quantum Social Science, Cambridge University Press
\bibitem{asano} Asano M., Khrennikov A., Ohya M., Tanaka Y., Yamato I. 2015 Quantum adaptivity in biology: From genetics to cognition, Springer
\bibitem{khrennikov_2} Khrennikov A., Haven E., Robinson T. 2017 Quantum methods in social sciences: a first course, World Scientific
\bibitem{bagarello_1} Bagarello F. 2012 Quantum dynamics for classical systems: with applications of the Number operator, John Wiley and Sons, Hoboken
\bibitem{bagarello_2} Bagarello F. 2019 Quantum Concepts in the Social, Ecological and Biological Sciences, Cambridge University Press
\bibitem{baaquie} Baaquie B.E. 2004 Quantum Finance, Cambridge University Press
\bibitem{busemeyer} Busemeyer J. R., Bruza P. D. 2012 Quantum models of cognition and decision, Cambridge University Press
\bibitem{galam} Galam S. 2012 Sociophysics, A Physicist’s Modeling of Psycho-political Phenomena, Springer

\bibitem{nava_traffic} Nava A., Giuliano D., Papa A., Rossi M. 2022 {\it SciPost Physics Core} {\bf 5} 022
\bibitem{lindblad_nrg} Nava A., Rossi M. and Giuliano D. 2021 {\it Phys. Rev.} B {\bf 103} 115139
\bibitem{petruccione} Breuer H.-P., Petruccione F. 2006 ``The Theory of Open Quantum Systems'', Oxford University, New York
\bibitem{shutz}  Schütz G. M., 1998 ``Integrable Stochastic Many-body Systems'', Berichte des Forschungszentrums Jülich
\bibitem{temme} Temme K., Wolf M. M., Verstraete F 2012 {\it New Journal of Physics} {\bf 14} 075004
\bibitem{paletta} De Leeuw M., Paletta C., Pozsgay B. 20212 {\it Phys. Rev. Lett.}  {\bf 126} 240403
\bibitem{2tasep} Bonnin P., Stansfield I., Romano M. C., Kern N 2022 {\it Phys. Rev.} E  {\bf 105} 034117
\bibitem{spin1} Tah R. 2020 {\it hal} 02909703 {\it DOI: 10.35543/osf.io/6ck3w}
\bibitem{ca_1} Weng W., Chen T., Yuan H., Fan W. 2006 {\it Phys. Rev.} E {\bf 74} 036102
\bibitem{ca_2} Feng S., Ding N., Chen T., Zhang H. 2013 {\it Physica A} {\bf 392} 2847 
\bibitem{ca_3} Liu M., Zeng W., Chen P., Wu X. 2017 {\it PLoS ONE} {\bf 12} e0180992
\bibitem{ca_4} Nava A., Papa A., Rossi M., Giuliano D. 2020 {\it Phys. Rev. Research} {\bf 2} 043379
\bibitem{fermi_1} Chamon C., Oshikawa M., Affleck I. 2003 {\it Phys. Rev. Lett.} {\bf 91} 206403
\bibitem{fermi_2} Oshikawa M., Chamon C., Affleck I. 2006 {\it J. Stat. Mech.} P02008
\bibitem{fermi_3} Guerci D., Nava A. 2021 {\it Physica E} {\bf 134} 114895
\bibitem{kondo_0} Tsvelik A. M. 2013 {\it Phys. Rev. Lett.} {\bf 110} 147202
\bibitem{kondo_1} Giuliano D., Lepori L., Nava A. 2020 {\it Phys. Rev.} B {\bf 101} 195140
\bibitem{kondo_2} Giuliano D., Nava A., Sodano P. 2020 {\it Nuclear Physics} B {\bf 960} 115192
\bibitem{kondo_3} Buccheri F., Nava A., Egger R., Sodano P., Giuliano D. 2022 {\it Phys. Rev.} B {\bf 105} L081403
\bibitem{fermi_5} Giuliano D., Nava A., Egger R., Sodano P., Buccheri F. 2022 {\it Phys. Rev.} B {\bf 105} 035419
\bibitem{ssh} Nava A., Campagnano G., Sodano P., Giuliano D. 2023 {\it Phys. Rev.} B {\bf 107} 035113


\bibitem{bagarello_3} Bagarello F., Di Salvo R., Gargano F., Oliveri F. 2017 Applied Mathematical Modelling {\bf 43} 15-32
\bibitem{graefe} Graefe E.-M., Honing M., Korsch H. J. 2010 J. Phys. A: Math. Theor. {\bf 43} 075306
\bibitem{bagarello_4} Bagarello F., Passante R., Trapani C. 2015 Non-Hermitian Hamiltonians in Quantum Physics, Springer International Publishing


\end{thebibliography}


\end{document}